\documentclass{article}
\usepackage{amsmath}
\usepackage{amsfonts}
\usepackage{amssymb}
\usepackage{graphicx}

\input{tcilatex}
\begin{document}

\title{Entropic Inference\thanks{%
Presented at MaxEnt 2010, the 30th International Workshop on Bayesian
Inference and Maximum Entropy Methods in Science and Engineering (July 4-9,
2010, Chamonix, France).}}
\author{Ariel Caticha \\
{\small Department of Physics, University at Albany-SUNY, }\\
{\small Albany, NY 12222, USA.}}
\date{}
\maketitle

\begin{abstract}
In this tutorial we review the essential arguments behing entropic
inference. We focus on the epistemological notion of information and its
relation to the Bayesian beliefs of rational agents. The problem of updating
from a prior to a posterior probability distribution is tackled through an
eliminative induction process that singles out the logarithmic relative
entropy as the unique tool for inference. The resulting method of Maximum
relative Entropy (ME), includes as special cases both MaxEnt and Bayes'
rule, and therefore unifies the two themes of these workshops -- the Maximum
Entropy and the Bayesian methods -- into a single general inference scheme.
\end{abstract}

\section{Introduction}

Our subject is inductive inference. Our goal in this tutorial paper is to
review the problem of updating from a prior probability distribution to a
posterior distribution when new information becomes available.

First we tackle the question of the nature of information itself: What is
information? It is clear that data \textquotedblleft
contains\textquotedblright\ or \textquotedblleft conveys\textquotedblright\
information, but what does this precisely mean? Is information physical? We
discuss how in a properly Bayesian framework one can usefully adopt a
concept of information that is more directly related to the epistemological
concerns of rational agents.

Then we turn to the actual methods to process information. We argue for the
uniqueness and universality of the Method of Maximum relative Entropy (ME)
and then we discuss its relation to Bayesian methods. At first sight
Bayesian and Maximum Entropy methods appear unrelated. Bayes' rule is the
natural way to update probabilities when the new information is in the form
of data. On the other hand, Jaynes' method of maximum entropy, MaxEnt, is
designed to handle information in the form of constraints \cite{Jaynes 57}.
An important question is whether they are compatible with each other. We
show that the ME method includes both MaxEnt and Bayesian methods as special
cases and allows us to extend them to situations that lie beyond the reach
of either of them individually.

Finally we explore an important extension of the ME method. The distribution
of maximum entropy has the highest probability of being the correct choice
of posterior, but how justified are we in ruling out those distributions
that do not maximize the entropy? The extended ME assigns a probability to
those other distributions and this has a wide variety of applications: it
provides a connection to the theory of large deviations, to f{}luctuation
theory, to entropic priors, and most recently to quantum mechanics. The
possibilities are endless.

We make no attempt to provide a review of the literature on entropic
inference. The following list, which ref{}lects only some contributions that
are directly related to the particular approach described in this tutorial,
is incomplete but might nevertheless be useful: Jaynes \cite{Jaynes 57},
Shore and Johnson \cite{ShoreJohnson 80}, Williams \cite{Williams 80},
Skilling \cite{Skilling 88}, Rodr\'{\i}guez \cite{Rodriguez 91}\cite%
{Rodriguez 98}, Giffin and Caticha \cite{Caticha 00}-\cite{Caticha 08}.

\section{What is information?}

The expression that systems \textquotedblleft carry\textquotedblright\ or
\textquotedblleft contain\textquotedblright\ information can perhaps be
traced to Shannon's theory of communication: a system is analogous to a
message. The system \textquotedblleft carries\textquotedblright\ information
about its own state and, in this sense, one can say that \emph{information
is physical}. Such physical\ information\ is directly associated to the
system. Our interest here is in an altogether different notion of
information which we might call epistemological and which is directly
associated to the beliefs of rational agents. Indeed, any fully Bayesian
theory of information requires an explicit account of how such
epistemological information is related to rational beliefs.

The need to update from one state of belief to another is driven by the
conviction that not all probability assignments are equally good; some
beliefs are preferable to others in the very pragmatic sense that they
enhance our chances to successfully navigate this world. The idea is that,
to the extent that we wish to be called rational, we will improve our
beliefs by revising them when new information becomes available: \emph{%
Information is what forces\ a change of rational beliefs.} Or, to put it
more explicitly:\emph{\ Information is a constraint on rational beliefs. }

This definition -- information is a constraint -- is sufficient for our
present purposes but would benefit from further elaboration. The definition
captures a notion of information that is directly related to changing our
minds: information is the driving force behind the process of learning. It
incorporates an important feature of rationality: being rational means
accepting that our beliefs must be constrained in very specific ways -- not
everything goes. But surely this is not enough: the indiscriminate
acceptance of any arbitrary constraint does not qualify as rational
behavior. To be rational an agent must exercise some judgement before
accepting a particular piece of information as a reliable basis for the
revision of its beliefs and this raises questions about what judgements
might be considered sound. Indeed, there is no implication that the
information must \emph{be true}; only that we \emph{accept it as true}.
False information is information too, at least as long as we are prepared to
accept it and allow it to affect our beliefs.

The paramount virtue of our definition is that it is useful. It allows
precise quantitative calculations even though the notion of an amount of
information, whether measured in bits or otherwise, is not introduced. By
information in its most general form, we just mean the set of constraints on
the family of acceptable posterior distributions and this is precisely the
kind of information the method of maximum entropy is designed to handle.

\section{Updating probabilities: the ME method}

The uncertainty about a variable $x\in \mathcal{X}$ (whether discrete or
continuous, in one or several dimensions) is described by a probability
distribution $q(x)$. Our goal is to design a method to update from a prior
distribution $q(x)$ to a posterior distribution $P(x)$ when new information
in the form of constraints becomes available. (The constraints can be given
in terms of expected values but this is not necessary. Other types of
constraints are allowed too; an example is appears in section 5.)

The problem is to select a distribution from among all those that satisfy
the constraints. The procedure is to rank the candidate distributions in
order of increasing preference \cite{Skilling 88}. It is clear that to
accomplish our goal the ranking must be transitive: if distribution $p_{1}$
is preferred over $p_{2}$, and $p_{2}$ is preferred over $p_{3}$, then $%
p_{1} $ is preferred over $p_{3}$. Such transitive rankings are implemented
by assigning to each $p(x)$ a real number $S[p]$ in such a way that if $%
p_{1} $ is preferred over $p_{2}$, then $S[p_{1}]>S[p_{2}]$. The selected
distribution $P$ (one or possibly many, for on the basis of the available
information we might have several equally preferred distributions) will be
that which maximizes the quantity $S[p]$, which we will henceforth call
entropy. We are thus led to a method of Maximum Entropy (ME) that involves
entropies that are real numbers and that are meant to be maximized. These
features are imposed by design; they are dictated by the function that the
ME method is being designed to perform and not by any objective properties
of the external world.

Next we must make a definite choice for the functional $S[p]$. Since the
purpose of the method is to update from priors to posteriors the ranking
scheme must depend on the particular prior $q$ and therefore the entropy $S$
must be a functional of both $p$ and $q$. Thus the entropy $S[p,q]$ produces
a ranking of the distributions $p$ \emph{relative} to the given prior $q$: $%
S[p,q]$ is the entropy of $p$ \emph{relative} to $q$. Accordingly $S[p,q]$
is commonly called \emph{relative entropy}, but since all entropies are
relative, even when relative to a uniform distribution, the modifier
`relative' is redundant and can be dropped.

The functional $S[p,q]$ is selected by a process of \emph{eliminative
induction}. The idea is simple: we start with a \emph{sufficiently broad}
family of candidates and identify a number of special cases for which we
know what the preferred distribution ought to be. Then we just eliminate all
those candidates that fail to provide the right update. As we shall see the
selection criteria adopted below are sufficiently constraining that there is
a single entropy functional $S[p,q]$ that survives the process of
elimination.

This approach has a number of virtues. First, to the extent that the
selection criteria are universally desirable, then the single surviving
entropy functional will be of universal applicability too. Second, the
reason why any entropy candidate is eliminated is quite explicit -- at least
one of the selection criteria is violated. Thus, the justification behind
the single surviving entropy is not that it leads to demonstrably correct
inferences, but rather, that other entropies are demonstrably wrong.

The selection criteria are chosen to ref{}lect the conviction that
information collected in the past and codified into the prior distribution
is valuable and should not be ignored. This attitude is very conservative:
the only aspects of one's beliefs that should be updated are those for which
new evidence has been supplied. Moreover, as we shall see below, the
selection criteria merely tell us what not to update, which has the virtue
of maximizing objectivity -- there are many ways to change something but
only one way to keep it the same. These ideas are summarized in the following

\textbf{Principle of Minimal Updating} (PMU): \emph{Beliefs must be revised
only to the extent required by the new information.}

Three selection criteria, a brief motivation for them, and their
consequences for the functional form of the entropy are listed below (proofs
and more details are given in \cite{Caticha 08}). The reason these criteria
are so constraining is that they refer to three infinitely large classes of
special cases where the desired update is known.

\textbf{Criterion 1: Locality}. \emph{Local information has local effects.}

\noindent If the information to be processed does not refer to an $x$ in a
particular subdomain $\mathcal{D\subset X}$ then the PMU requires that we do
not change our minds about $x\in \mathcal{D}$. More precisely, we require
that the prior conditioned on $\mathcal{D}$ is not updated. The selected
posterior is such that $P(x|\mathcal{D})=q(x|\mathcal{D})$. Dropping
additive terms and multiplicative factors that do not affect the overall
ranking, the surviving entropy functionals are of the form 
\begin{equation}
S[p,q]=\int dx\,F\left( p(x),q(x),x\right) \ ,  \label{axiom1}
\end{equation}%
where $F$ is some unknown function and by $\tint dx$ we mean a discrete sum
or continuous integral (possibly over several dimensions) as the case might
require.

\textbf{Criterion 2: Coordinate invariance.} \emph{The system of coordinates
carries no information. }

\noindent The points $x$ can be labeled in different ways using different
coordinate systems but this should not affect the ranking of the
distributions. The consequence of criterion 2 is that the surviving
entropies can be written as 
\begin{equation}
S[p,q]=\int dx\,m(x)\Phi \left( \frac{p(x)}{m(x)},\frac{q(x)}{m(x)}\right) ~,
\label{axiom2}
\end{equation}%
where $m(x)$ is a probability density, which implies that $dx\,m(x)$, $%
p(x)/m(x)$, and $q(x)/m(x)$ are coordinate invariants. (Again, additive
terms and multiplicative factors that do not affect the overall ranking have
been dropped.) We see that the single unknown function $F$ in (\ref{axiom1})
with three arguments has been replaced by two unknown functions. One is the
density $m(x)$, and the other is a function $\Phi $ with two arguments. The
density $m(x)$ is determined by invoking the locality criterion once again.

\textbf{Criterion 1 (a special case): }\emph{When there is no new
information there is no reason to change one's mind. }

\noindent When no new information is available the domain $\mathcal{D}$ in
criterion 1 coincides with the whole space $\mathcal{X}$. The conditional
probabilities $q(x|\mathcal{D})=q(x|\mathcal{X})=q(x)$ should not be updated
and the selected posterior coincides with the prior, $P(x)=q(x)$. The
consequence is that up to normalization the unknown $m(x)$ must be the prior
distribution $q(x)$. The entropy is now restricted to functionals of the
form 
\begin{equation}
S[p,q]=\int dx\,q(x)\Phi \left( \frac{p(x)}{q(x)}\right) ~.
\end{equation}

\textbf{Criterion 3:\ Independence}. \emph{When systems are known\ to be
independent it should not matter whether they are treated separately or
jointly. }

\noindent The preservation of independence is a particularly important
concern for science because without it science is not possible. The reason
is that in any inference it is assumed that the universe is partitioned into
the system of interest and other systems that constitute the rest of the
universe. What is important about those other systems is precisely that they
can be ignored -- whether they are included in the analysis or not should
make no difference. If they did matter they should have been incorporated as
part of the system of interest in the first place.

It is crucial that Criterion 3 be applied to all independent systems whether
they are identical or not, whether just two or many, or even infinitely
many. This criterion is sufficiently constraining that (up to additive terms
and multiplicative factors that do not affect the overall ranking scheme)
there is a single surviving entropy functional given by the usual
logarithmic relative entropy \cite{Caticha 08}, 
\begin{equation}
S[p,q]=-\int dx\,p(x)\log \frac{p(x)}{q(x)}  \label{S}
\end{equation}

\noindent These results are summarized as follows:

\textbf{The ME method:} \emph{The objective is to update from a prior
distribution }$q$\emph{\ to a posterior distribution }$P$\emph{\ given the
information that the posterior lies within a certain family of distributions 
}$p$\emph{. The selected posterior }$P$\emph{\ is that which maximizes the
entropy }$S[p,q]$\emph{. Since prior information is valuable the functional }%
$S[p,q]$\emph{\ is chosen so that beliefs are updated only to the minimal
extent required by the new information. No interpretation for }$S[p,q]$\emph{%
\ is given and none is needed.}

\section{Bayes' rule and its generalizations}

Bayes' rule is used to make inferences about one or several quantities $%
\theta \in \Theta $ on the basis of information in the form of data $x\in 
\mathcal{X}$. More specifically, the problem is to update our beliefs about $%
\theta $ on the basis of three pieces of information: (1) the prior
information codified into a prior distribution $q(\theta )$; (2) the data $%
x\in \mathcal{X}$ (obtained in one or many experiments); and (3) the known
relation between $\theta $ and $x$ given by the model as defined by the
sampling distribution or likelihood function, $q(x|\theta )$. The updating
consists of replacing the \emph{prior} probability distribution $q(\theta )$
by a \emph{posterior} distribution $P(\theta )$ that applies after the data
has been processed.

\noindent \textbf{Remark:} We emphasize that the information about how $x$
is related to $\theta $ is contained in the \emph{functional form} of the
distribution $q(\cdot |\theta )$ which is completely unrelated to the actual
values of the observed data.

The insight that will allow Bayes' rule to be smoothly incorporated into the
entropic inference framework \cite{Williams 80}\cite{Caticha Giffin 06} is
that the relevant universe of discourse is not $\Theta $ but the product
space $\Theta \times \mathcal{X}$ \cite{Rodriguez 91}\cite{Rodriguez 98}. We
deal with joint distributions and the relevant \emph{joint prior} is $%
q(x,\theta )=q(\theta )q(x|\theta )$.

\noindent \textbf{Remark:} Bayes' rule is usually written in the form 
\begin{equation}
q(\theta |x)=q(\theta )\frac{q(x|\theta )}{q(x)}~,
\end{equation}%
and called Bayes' theorem. This formula is a restatement of the product
rule. It is valid for any value of $x$ whether it coincides with the
observed data or not and therefore it is a simple consequence of the \emph{%
internal} consistency of the \emph{prior} beliefs. Within the framework of
entropic inference the left hand side is not a \emph{posterior} but rather a 
\emph{prior} probability -- it is the prior probability of $\theta $
conditional on $x$.

Next we collect data and the observed values turn out to be $x^{\prime }$.
This constrains the posterior to the family of distributions $p(x,\theta )$
defined by%
\begin{equation}
p(x)=\tint d\theta \,p(\theta ,x)=\delta (x-x^{\prime })~.
\label{data constraint a}
\end{equation}%
This data information is not, however, sufficient to determine the joint
distribution 
\begin{equation}
p(x,\theta )=p(x)p(\theta |x)=\delta (x-x^{\prime })p(\theta |x^{\prime })~.
\end{equation}%
Any choice of $p(\theta |x^{\prime })$ is in principle possible. Within the
framework of entropic inference (see \cite{Caticha Giffin 06}) the joint
posterior $P(x,\theta )$ is the minimal update from the prior $q(x,\theta )$
that agrees with the data constraint. To find it maximize the entropy, 
\begin{equation}
S[p,q]=-\tint dxd\theta ~p(x,\theta )\log \frac{p(x,\theta )}{q(x,\theta )}%
~,~  \label{entropy}
\end{equation}%
subject to the \emph{infinite} number of constraints given by eq. (\ref{data
constraint a}). Note that there is one constraint for each value of $x$. The
corresponding Lagrange multipliers are denoted $\lambda (x)$. Maximizing (%
\ref{entropy}) subject to (\ref{data constraint a}) and normalization, 
\begin{equation}
\delta \left\{ S+\alpha \left[ \tint dxd\theta ~p(x,\theta )-1\right] +\tint
dx\,\lambda (x)\left[ \tint d\theta ~p(x,\theta )-\delta (x-x^{\prime })%
\right] \right\} =0~,
\end{equation}%
yields 
\begin{equation}
P(x,\theta )=q(x,\theta )\,\frac{e^{\lambda (x)}}{Z}~,
\end{equation}%
where $Z$ is a normalization constant, and $\lambda (x)$ is determined from (%
\ref{data constraint a}), 
\begin{equation}
\tint d\theta ~q(x,\theta )\frac{\,e^{\lambda (x)}}{Z}=q(x)\frac{%
\,e^{\lambda (x)}}{Z}=\delta (x-x^{\prime })~,
\end{equation}%
so that the joint posterior is 
\begin{equation}
P(x,\theta )=q(x,\theta )\frac{\,\delta (x-x^{\prime })}{q(x)}=\delta
(x-x^{\prime })q(\theta |x)~.
\end{equation}%
The corresponding marginal posterior probability $P(\theta )$ is 
\begin{equation}
P(\theta )=\tint dx\,P(\theta ,x)=q(\theta |x^{\prime })=q(\theta )\frac{%
q(x^{\prime }|\theta )}{q(x^{\prime })}~,  \label{Bayes rule}
\end{equation}%
which coincides with Bayes' rule. This is intuitively reasonable: we \emph{%
maintain} those beliefs about $\theta $ that are consistent with the data
values $x^{\prime }$ that turned out to be true. Data values that were not
observed are discarded because they are now known to be false. The extension
to repeatable independent experiments is straightforward \cite{Caticha 08}.

Next I give a couple of very simple examples that show how entropic methods
allow generalizations of Bayes' rule.

\noindent \textbf{Example 1.--} \textbf{Jeffrey's rule.} As before, the
prior information consists of our prior knowledge about $\theta $ given by
the distribution $q(\theta )$ and the relation between $x$ and $\theta $ is
given by the likelihood $q(x|\theta )$. But now the information about $x$ is
limited because the data is uncertain. The marginal posterior $p(x)$ is no
longer a sharp delta function but some other known distribution, $%
p(x)=P_{D}(x)$. This is still an infinite number of constraints 
\begin{equation}
p(x)=\tint d\theta \,p(\theta ,x)=P_{D}(x)~.  \label{data constraint b}
\end{equation}%
Maximizing (\ref{entropy}) subject to (\ref{data constraint b}) and
normalization, leads to 
\begin{equation}
P(x,\theta )=P_{D}(x)q(\theta |x)~.
\end{equation}%
The corresponding marginal posterior, 
\begin{equation}
P(\theta )=\tint dx\,P_{D}(x)q(\theta |x)=q(\theta )\tint dx\,P_{D}(x)\frac{%
q(x|\theta )}{q(x)}~,  \label{Jeffrey}
\end{equation}%
is known as Jeffrey's rule. In the limit when the data are sharply
determined $P_{D}(x)=\delta (x-x^{\prime })$ the posterior reproduces Bayes'
rule (\ref{Bayes rule}).

\noindent \textbf{Example 2.--} \textbf{Unknown likelihood.} The following
example derives and generalizes Zellner's Bayesian Method of Moments \cite%
{Zellner 97}. Usually the relation between $x$ and $\theta $ is given by a
known likelihood function $q(x|\theta )$ but suppose this relation is not
known. This is the case when the joint prior is so ignorant that information
about $x$ tells us nothing about $\theta $ and vise versa; such a prior
treats $x$ and $\theta $ as statistically independent, $q(x,\theta
)=q(x)q(\theta )$. Since we have no likelihood function the information
about the relation between $\theta $ and the data $x$ must be supplied
elsewhere. One possibility is through a constraint. Suppose that in addition
to normalization and the uncertain data constraint, eq.(\ref{data constraint
b}), we also know that the expected value over $\theta $ of a function $%
f(x,\theta )$ is 
\begin{equation}
\langle f\rangle _{x}=\tint d\theta \,p(\theta |x)f(x,\theta )=F(x)~.
\label{data constraint c}
\end{equation}%
We seek a posterior $P(x,\theta )$ that maximizes (\ref{entropy}).
Introducing Lagrange multipliers $\alpha $, $\lambda (x)$, and $\gamma (x)$, 
\begin{eqnarray}
0 &=&\delta \left\{ S+\alpha \left[ \tint dxd\theta ~p(x,\theta )-1\right]
+\tint dx\,\lambda (x)\left[ \tint d\theta ~p(x,\theta )-P_{D}(x)\right]
\right. \\
&&\left. +\tint dx\,\gamma (x)\left[ \tint d\theta ~p(x,\theta )f(x,\theta
)-P_{D}(x)F(x)\right] \right\} ~,
\end{eqnarray}%
the variation over $p(x,\theta )$ yields 
\begin{equation}
P(x,\theta )=\frac{1}{\zeta }q(x)q(\theta )\,e^{\lambda (x)+\gamma
(x)f(x,\theta )}~,
\end{equation}%
where $\zeta $ is a normalization constant. The multiplier $\lambda (x)$ is
determined from (\ref{data constraint a}), 
\begin{equation}
P(x)=\tint d\theta \,P(\theta ,x)=\frac{1}{\zeta }q(x)e^{\lambda (x)}\tint
d\theta \,q(\theta )\,e^{\gamma (x)f(x,\theta )}=P_{D}(x)
\end{equation}%
then, 
\begin{equation}
P(x,\theta )=P_{D}(x)\frac{q(\theta )\,e^{\gamma (x)f(x,\theta )}}{\tint
d\theta ^{\prime }\,q(\theta ^{\prime })\,e^{\gamma (x)f(x,\theta ^{\prime
})}}
\end{equation}%
so that 
\begin{equation}
P(\theta |x)=\frac{P(x,\theta )}{P(x)}=\frac{q(\theta )\,e^{\gamma
(x)f(x,\theta )}}{Z(x)}\quad \text{with}\quad Z(x)=\tint d\theta ^{\prime
}\,q(\theta ^{\prime })\,e^{\gamma (x)f(x,\theta ^{\prime })}
\end{equation}%
The multiplier $\gamma (x)$ is determined from (\ref{data constraint c}) 
\begin{equation}
\frac{1}{Z(x)}\frac{\partial Z(x)}{\partial \gamma (x)}=F(x)~.
\end{equation}%
The corresponding marginal posterior is 
\begin{equation}
P(\theta )=\tint dx\,P_{D}(x)P(\theta |x)=q(\theta )\tint dx\,P_{D}(x)\frac{%
\,e^{\gamma (x)f(x,\theta )}}{Z(x)}~.
\end{equation}%
In the limit when the data are sharply determined $P_{D}(x)=\delta
(x-x^{\prime })$ the posterior takes the form of Bayes theorem, 
\begin{equation}
P(\theta )=q(\theta )\,\frac{e^{\gamma (x^{\prime })f(x^{\prime },\theta )}}{%
Z(x^{\prime })}~,
\end{equation}%
where up to a normalization factor $e^{\gamma (x^{\prime })f(x^{\prime
},\theta )}$ plays the role of the likelihood and the normalization constant 
$Z$ plays the role of the evidence.

In conclusion, these examples demonstrate that the method of maximum entropy
can fully reproduce the results obtained by the standard Bayesian methods
and allows us to extend them to situations that lie beyond their reach such
as when the likelihood function is not known. Other such examples are given
in \cite{Giffin Caticha 07} and \cite{Caticha 08}.

\section{Deviations from maximum entropy}

To complete the design of the ME method we must address one last issue. Once
we have decided that the distribution that maximizes entropy is to be
preferred over all others we ask: to what extent are the other distributions
ruled out? The discussion below follows \cite{Caticha 03}\cite{Caticha 08}.

The original problem was to update from a prior $q(x)$ given constraints
that define the space $\Theta $ of acceptable distributions. We assume that
these distributions, that is, the \textquotedblleft
points\textquotedblright\ in the space $\Theta $, can be labelled by
coordinates $\theta $. Thus, $\Theta $ is a statistical manifold and its
points can be written as $p(x|\theta )$. Maximizing $S[p,q]$ over all the $%
p(x|\theta )$ in $\Theta $ leads to the preferred distribution, say $%
p(x|\theta _{0})$.

The question about the extent that distributions with $\theta \neq \theta
_{0}$ are ruled out is a question about the probability of various values of 
$\theta $: to what extent do we believe that the selected value should lie
within any particular range $d\theta $? Thus we are not just concerned with
the probability of $x$, but with the joint distribution $p(x,\theta )$. To
assign $p(x,\theta )$ we apply the same ME method but in the larger joint
space: maximize the joint entropy%
\begin{equation}
\mathcal{S}[p,q]=-\int dx\,d\theta \,p(x,\theta )\log \frac{p(x,\theta )}{%
q(x,\theta )}~,  \label{S joint a}
\end{equation}%
for a suitable prior $q(x,\theta )$ and under the appropriate constraints.

Choosing a prior is always tricky because it represents what we knew \emph{%
before }the relevant new information became available. We want to represent
a state of extreme ignorance: the precise relation between $\theta $s and $x$%
s is not (yet) known and therefore $q(x,\theta )$ is a product, $q(x,\theta
)=q(x)q(\theta )$, so that knowing $x$ tells us nothing about $\theta $ and
vice versa. For $q(x)$ we retain the prior used in the original problem
where we updated from $q(x)$ to $p(x|\theta _{0})$.

For $q(\theta )$ we plead ignorance once again and choose a uniform
distribution. This is somewhat trickier than may seem at first sight because 
\emph{uniform }does not mean \emph{constant}. The uniform distribution
assigns equal probabilities to equal volumes in $\Theta $ and does not
depend on the particular choice of coordinates. (A constant distribution, on
the other hand, depends on the choice of coordinates: a distribution that is
constant in one frame coordinate will not be constant in another.) This
requires a well-defined notion of volume. Fortunately, the statistical
manifold $\Theta $ is a metric space:\ \emph{there is a single unique
geometry} that properly takes into account the fact that the points in $%
\Theta $ are not structureless points but are actual probability
distributions. This is given by the Fisher-Rao information metric, $%
g_{ij}(\theta )$\emph{\ }\cite{Amari 85}\cite{Caticha 08}\emph{.} The
corresponding volume elements are given by $g^{1/2}(\theta )d^{n}\theta $,
where $g=\det g_{ij}$. Therefore the uniform (unnormalized) prior is $%
q(\theta )=g^{1/2}(\theta )$ and the joint prior is $q(x,\theta
)=g^{1/2}(\theta )q(x)$.

The crucial constraint on the joint distributions $p(x,\theta )=p(\theta
)p(x|\theta )$ specifies the conditional distributions $p(x|\theta )$. This
amounts to selecting the particular space $\Theta $ under consideration.

The preferred joint distribution $P(x,\theta )$ is that which maximizes the
joint entropy $\mathcal{S}[p,q]$ over all normalized distributions of the
form $p(x,\theta )=p(\theta )p(x|\theta )$ where we vary with respect to $%
p(\theta )$ and restrict to $p(x|\theta )\in \Theta $. It is convenient to
rewrite (\ref{S joint a}) as 
\begin{equation}
\mathcal{S}[p,q]=-\int \,d\theta \,p(\theta )\log \frac{p(\theta )}{%
g^{1/2}(\theta )}+\int d\theta \,p(\theta )S(\theta ),  \label{S joint b}
\end{equation}%
where 
\begin{equation}
S(\theta )=-\int \,dx\,p(x|\theta )\log \frac{p(x|\theta )}{q(x)}.
\label{Stheta}
\end{equation}%
The result is the probability that $\theta $ lies within a small volume $%
g^{1/2}(\theta )d^{n}\theta $, 
\begin{equation}
P(\theta )d^{n}\theta =\frac{1}{\zeta }\,\,e^{S(\theta )}g^{1/2}(\theta
)d^{n}\theta \quad \text{with}\quad \zeta =\int d^{n}\theta \,g^{1/2}(\theta
)\,e^{S(\theta )}\,.  \label{main}
\end{equation}%
The preferred value of $\theta $ is that $\theta _{0}$ which maximizes the
entropy $S(\theta )$, eq.(\ref{Stheta}), because this maximizes the scalar
probability density $\exp S(\theta )$. But it also tells us the degree to
which values of $\theta $ away from the maximum are ruled out.

One of the limitations of the standard MaxEnt method is that it selects a
single \textquotedblleft posterior\textquotedblright\ $p(x|\theta _{0})$ and
all other distributions are strictly ruled out. The result (\ref{main})
overcomes this limitation and finds many applications. For example, it
extends the Einstein theory of thermodynamic f{}luctuations beyond the
regime of small f{}luctuations; it provides a bridge to the theory of large
deviations; and, suitably adapted for Bayesian data analysis, it leads to
the notion of entropic priors.

\section{Conclusions}

Any Bayesian account of the notion of information cannot ignore the fact
that Bayesians are concerned with the beliefs of rational agents. The
relation between information and beliefs must be clearly spelled out. The
definition we have proposed -- that information is that which constrains
rational beliefs and therefore forces the agent to change its mind -- is
convenient for two reasons. First, the information/belief relation is
explicit, and second, the definition is ideally suited for quantitative
manipulation using the ME\ method.

The main conclusion is that the logarithmic relative entropy is the only
candidate for a general method for updating probabilities -- the ME\ method
-- and this includes both MaxEnt and Bayes' rule as special cases; it
unifies them into a single theory of inductive inference and allows new
applications. Indeed, much as the old MaxEnt method provided the foundation
for statistical mechanics, recent work suggests that the extended ME method
provides an entropic foundation for quantum mechanics.

\noindent \textbf{Acknowledgements:} I would like to acknowledge valuable
discussions with C. Cafaro, N. Caticha, A. Giffin, K. Knuth, C. Rodr\'{\i}%
guez, J. Skilling, and C.-Y. Tseng.

\end{document}